\documentclass[preprint,12pt]{elsarticle}

\usepackage{amsmath,amssymb}
\usepackage{color}

\journal{}

\begin{document}

\begin{frontmatter}



\title{Self-reinforcing feedback loop in financial markets with coupling of market impact and momentum traders}


\author{Li-Xin Zhong$^{a}$}\ead{zlxxwj@163.com}
\author {Wen-Juan Xu$^b$}
\author {Rong-Da Chen$^a$}
\author{Chen-Yang Zhong$^c$}
\author {Tian Qiu$^d$}
\author {Fei Ren$^e$}
\author {Yun-Xing He$^a$}

\address[label1]{School of Finance and Coordinated Innovation Center of Wealth Management and Quantitative Investment, Zhejiang University of Finance and Economics, Hangzhou, 310018, China}
\address[label2]{School of Law, Zhejiang University of Finance and Economics, Hangzhou, 310018, China}
\address[label3]{Department of Statistics, Stanford University, Stanford, CA 94305-4065, USA}
\address[label4]{School of Information Engineering, Nanchang Hangkong University, Nanchang, 330063, China}
\address[label5]{School of Business and Research Center for Econophysics, East China University of Science and Technology, Shanghai, 200237, China}

\begin{abstract}
By incorporating market impact and momentum traders into an agent-based model, we investigate the conditions for the occurrence of self-reinforcing feedback loops and the coevolutionary mechanism of prices and strategies. For low market impact, the price fluctuations are originally large. The existence of momentum traders has little impact on the change of price fluctuations but destroys the equilibrium between the trend-following and trend-rejecting strategies. The trend-following herd behaviors become dominant. A self-reinforcing feedback loop exists. For high market impact, the existence of momentum traders leads to an increase in price fluctuations. The trend-following strategies of rational individuals are suppressed while the trend-following strategies of momentum traders are promoted. The crowd-anticrowd behaviors become dominant. A negative feedback loop exists. A theoretical analysis indicates that, for low market impact, the majority effect is beneficial for the trend-followers to earn more, which in turn promotes the trend-following strategies. For high market impact, the minority effect causes the trend-followers to suffer great losses, which in turn suppresses the trend-following strategies.

\end{abstract}


\end{frontmatter}


\section{Introduction}
\label{sec:introduction}
Bubbles and crashes occasionally occur in real estate and financial markets, which can swallow years of individual savings and trigger economic recessions \cite{johansen1, roehner1}. Empirical analyses show that positive feedbacks give a driving force for the growth of speculative bubbles in financial markets and cause unsustainable growth of the housing prices \cite{sornette1, zhou1, zhou2, zhou3}.  The individuals who are interested in fast gains can promote the self-reinforcing imitation, which finally leads to the speculative herding and the short-term price momentum and long-term price reversal \cite{teplova1, gutierrez1, eguiluz1, chen20}.

In the last two decades, the world has witnessed the booming growth of bounded rationality and behavioural finance theory (BFT) \cite{bloomfield1, barberis1, takahashi1}, which is one of the strongest challenges to the efficient market hypothesis (EMH). Different from the EMH, the BFT attributes the irrational behaviors in financial markets to cognitive biases. Bloomfield et al. have investigated rational cognition and cognitive biases on price movement. The price fluctuations were found to be closely related to the degree of risk preference \cite{bloomfield1, barberis1}. Kukacka et al. have examined the direct impact of herding, overconfidence, and market sentiment in financial market \cite{kukacka1}. The behavioural patterns were found to be quite important for the evolution of prices. The stock prices are quite possible to overreact or underreact to the news, which can be explained by the behavioral biases like over-confidence or risk-aversion \cite{gutierrez1, zhong1, zhong2}. 

In exploring the evolutionary dynamics in financial markets, a variety of agent-based models have been borrowed to mimic the stock price movement \cite{wei1, chen1, chen2, li1, lo1, hadzibeganovic20, han20}. The most relevant models are the Ising models, the herding models and the minority game models \cite{sornette2, zhou4, sornette3, eguiluz1, chen20, yeung1, barato1, challet1, johnson1, johnson2}, which can reproduce some of the stylized facts found in realistic markets \cite{mike1, gu1, gu2}. By incorporating herding, external news and private information into a dynamical Ising-like model, the stylized facts of financial markets have been reproduced \cite{zhou4}. The seminal work of Mike and Farmer proposes an empirical behavioral model based on the empirical regularities at the trade level \cite{mike1}, which has been further improved by including long-range correlations in relative prices and price limit rules \cite{gu1, gu2}. These models have been applied to study the micro mechanisms causing macro properties \cite{gu3, meng1, zhou5, chen20}. A behavioral model has been proposed to mimick the stock price formation, it can successfully reproduce the distribution of returns, the well-known power-law tails, some other important stylized facts \cite{gu3}.

Until today, in relation to the role of behavioural biases in investment, some empirical findings indicate that they help the investors earn more while other empirical findings indicate that they cause the investors to suffer a loss\cite{roehner1,sornette1}. Both the efficient market hypothesis and the behavioral finance theory can explain part of these phenomena. Whether the contradictory scenarios can coexist or not and how the rational and irrational behaviors can be coupled together are still open questions.

Following the work done by bloomfield and Yeung et al.\cite{bloomfield1,yeung1}, in the present work, we incorporate market impact and momentum traders into an agent-based model. The conditions for the occurrence of self-reinforcing feedback loops and the coevolutionary mechanism of stock prices and individual strategies are extensively investigated. We have the following three main findings.

(1) The effects of momentum traders on price fluctuations are determined by the market impact. As the market impact is low, the price fluctuations are quite large even if there are no momentum traders. The existence of momentum traders has little impact on the change of price fluctuations. As the market impact is high, the price fluctuations are quite small when there are no momentum traders. The existence of momentum traders leads to an increase in the price fluctuations.

(2) The self-reinforcing feedback loop only exists within the range where the market impact is low. For low market impact, the trend-following strategies of momentum traders are imitated by rational individuals. The trend-following herd behavior occurs easily. For high market impact, the trend-following strategies of momentum traders are rejected by rational individuals. The trend-rejecting strategies of rational individuals coexist with the trend-following strategies of momentum traders.

(3) A theoretical analysis indicates that the trend-following or trend-rejecting behavior results from the gains or losses in the investment. As the market impact is low, the individuals adopting the trend-following strategy are easy to be in the majority side and gain more, which in turn attracts more rational individuals to adopt the trend-following strategy. As the market impact is high, the trend-rejecting strategy accompanied by the trend-following strategy are easy to be in the minority side, which in turn leads to the coexistence of the trend-rejecting strategies of rational individuals and the trend-following strategies of momentum traders.

The paper is organized as follows. In section 2, the evolutionary minority game with market impact and momentum traders is introduced. In section 3, the numerical results are presented. In section 4, a theoretical analysis is given. In section 5, some valuable conclusions are drawn.

\section{The model}
\label{sec:model}
The price formation is related to the dependence of return on trade volume. The price impact against the transaction size has been investigated in refs. \cite{lillo1, lim1, zhou6, zhou7, xu1, pham1}. A uniform price-impact curve suitable for different markets has been investigated,  the power-law dependence of the return on the size has been found \cite{lillo1, lim1}. The empirically estimated exponent ranges from $\frac{1}{3}$ to $\frac{2}{3}  $\cite{zhou6, zhou7, xu1, pham1}.

Following the work done in refs. \cite{zhou7, xu1}, in the present model, we introduce a variable $\xi$ as the power-law price impact. The time-dependent evolution of stock prices is satisfied with the equation

\begin{equation}
P(t+1)=P(t)e^{\frac{\eta sign[A(t)] {\mid A(t)\mid^\xi}}{N}},
\end{equation}
in which $A(t)=\sum_{i=1}^{N}a_i(t)$ and $a_i(t)$ is an individual $i's$  decision, buying (+1), selling (-1) or taking no action (0). $\xi$ ranges from 0 to 1 and $\eta$ is a pregiven constant. If there are more buyers than sellers, $A(t)$ is greater than zero and the price increases. If there are more sellers than buyers, $A(t)$ is less than zero and the price decreases.

Consider a stock market with $\gamma N$ momentum traders and $(1-\gamma) N$ rational individuals, $\gamma\in[0,1]$. For a rational individual $i$, he makes his decision according to his strategy $g_i\in[0,1]$. Facing a m-bit long history indicating the rise and fall of the latest prices, he follows the historical prediction with probability $g_i$ and rejects the historical prediction with probability $1-g_i$. For example, facing a price history $\uparrow\uparrow\uparrow\uparrow\uparrow$, if the historical prediction is $\downarrow$, rational individual $i$ sells the stock with probability $g_i$ on condition that he has a stock in his hand and buys the stock with probability $1-g_i$ on condition that he has no stock in his hand. Or else, he takes no action. For a momentum trader $j$, he acts as a trend-follower when some typical history occurs. For example, facing a continuous rise of the stock prices, $\uparrow\uparrow\uparrow$ or $\uparrow\uparrow\uparrow\uparrow$, on condition that momentum trader $j$ has no stock in his hand, he buys the stock no matter what the historical prediction is. Facing a continuous fall of the stock prices, $\downarrow\downarrow\downarrow$ or $\downarrow\downarrow\downarrow\downarrow$, on condition that he has a stock in his hand, he sells the stock no matter what the historical prediction is. Or else, he acts as a rational individual.

Strategy $g$ evolves according to the strategy score $D$ and the updating threshold $D_{th}$,

\begin{equation}
D=\sum_{t=t_{update}}^{t_{latest}}[P_{tr}(t_{sell})-P_{tr}(t_{buy})],
\end{equation}
in which $D$ is the accumulated wealth received after the new strategy has been adopted and $P_{tr}$ is the transaction price. On condition that $D_i<D_{th}$, individual $i$ modifies his strategy $g_i$ and gets a new strategy randomly from the range of $\lbrack g_i-\frac{\lambda}{2},g_i+\frac{\lambda}{2}\rbrack$, $\lambda\in[0,1]$.

The market impact is incorporated into the transaction price $P_{tr}$ with parameter $\beta$ \cite{yeung1},

\begin{equation}
P_{tr}(t)= (1-\beta)P(t)+ \beta P(t+1),
\end{equation}
in which $0\leq \beta \leq 1$. For $\beta=0$, $P_{tr}(t)= P(t)$, the individuals make a transaction according to the latest price. For $\beta=1$, $P_{tr}(t)= P(t+1)$, the individuals make a transaction according to the next price.

The wealth of an individual $i$ is accumulated from the beginning of the game,

\begin{equation}
W_i=\sum_{t=1}^{t_{max}}\Delta W_i(t),
\end{equation}
in which $\Delta W=P_{tr}(t_{sell})-P_{tr}(t_{buy})$. The averaged wealth of the population is

\begin{equation}
\bar W=\frac{1}{N} \sum_{i=1}^{N}W_i.
\end{equation}

The predictability is used to measure the mean value of the change in price as some typical history occurs, which is defined as \cite{yeung1}

\begin{equation}
H=\sum_{\chi}\rho{(\chi)}\langle\Delta P\vert\chi\rangle^2,
\end{equation}
in which $\rho{(\chi)}$ is the probability that history $\chi$ occurs and $\langle\Delta P\vert\chi\rangle$ is the mean value of the change in price on condition that history $\chi$ occurs.

The probability distribution is often used to find the empirical characteristics of a series of data. Given a time-dependent series of price $P(t)$, $t$=1, 2, 3, ... , $t_{max}$, $t_{max}+1$. The return is $r(t)=ln P(t+1)-ln P(t)$. The normalized return is $R(t)=r(t)/\vert r\vert_{max}$, in which $\vert r\vert_{max}$ is the maximal value of absolute return. In real financial markets, the positive and negative tails of the probability distribution of stock returns is about 3 \cite{plerou1}.

The detrended fluctuation analysis (DFA) method is often used to quantify the long-range correlations in signals \cite{peng1, peng2}. In financial markets, the DFA method is often applied to find the empirical characteristics of stock prices \cite{liu1, chen10, qiu1}. Given a time-dependent series of variable $x(t)$, $t$=1, 2, 3, ... , $t_{max}$, $t_{max}+1$. It is integrated as $y(k)=\sum_{i=1}^{k}[x(i)-x_{ave}]$, in which $x(i)$ is the $ith$ value and $x_{ave}$ is the average value. The integrated series is divided into $n$ boxes and the length of each box is $S$. The detrended series of $y(k)$ is $y(k)-y_s(k)$, in which $y_s(k)$ is the local trend of the detrended series in each box. The root-mean-square of $y(k)$ is $F(S)=\sqrt{\frac{\sum_{k=1}^{S}[y(k)-y_s(k)]^2}{S}}$. The relationship between $F(S)$ and S, $F(S)\sim S^h$, can be understood as follows. For $h<0.5$, the series is satisfied with a short-term correlation. For $h=0.5$, the series is a random walk. For $h>0.5$, the series is satisfied with a long-range correlation.

In real financial markets, the trend-following momentum traders are found here and there. For example, during a bull market, some investors rush into the market no matter what a fundamental analysis is. During a bear market, some investors run away from the market no matter what a rational analyst tells them to do. The momentum trading strategy is frequently used in security investment \cite{shi1, shi2}. How the existence of momentum traders disturb the originally rational population should be especially examined.

\section{Simulation results and discussions}
\label{sec:results}
\subsection{\label{subsec:levelA} Stylized facts of stock returns and stock return volatilities}

Some empirical studies on real market data have shown that the stock price movement exhibits some stylized facts such as the fat-tails of return distribution, no memory in return and long-range autocorrelation in return volatilities \cite{li1, gu1, gu3, gopikrishnan1}. In the following, we firstly examine whether the proposed model can reproduce some of the well-known stylized facts found in real financial markets.

\begin{figure}
\includegraphics[width=10cm]{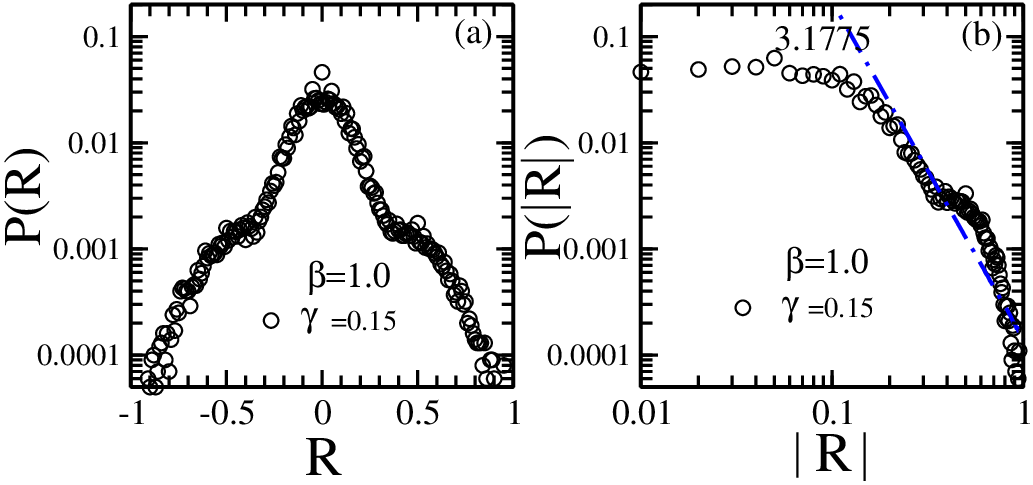}
\caption{\label{fig:epsart}(a) The distribution $P(R)$ of normalized stock returns; (b) the distribution $P(\mid R\mid)$ of absolute normalized stock returns for market impact $\beta=1.0$ and the ratio of momentum traders $\gamma$= 0.15 (circles). Other parameters are: the total number of individuals $N=1001$, the memory size $m=5$, the updating threshold $D_{th}=0$, the drift of strategy $\lambda=0.1$, the exponent $\xi=1$, the constant $\eta=10$. Final results are averaged over 10 runs and $10^4$ times after $10^5$ relaxation times in each run.}
\end{figure}

In fig. 1 (a) we plot the distribution $P(R)$ of normalized stock returns. It is noted that, compared with a normal distribution, $P(R)$ has a relatively fat tail. In fig. 1 (b) we plot the distribution $P(\mid R\mid)$ of simulation data and draw a best fit line with double logarithmic approximation. A power-law distribution with an exponent $\alpha\sim 3.1775$ is observed. Such a result is in accordance with the empirical findings in refs. \cite{gu3, gopikrishnan1}, where both the positive and negative tails are about $\alpha\sim 3$.

\begin{figure}
\includegraphics[width=10cm]{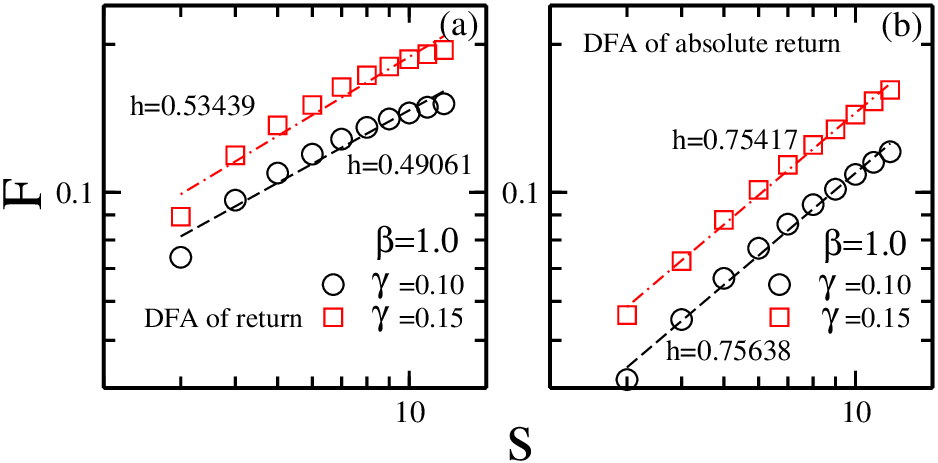}
\caption{\label{fig:epsart}(a) DFA of stock returns; (b) DFA of absolute stock returns for the market impact $\beta=1.0$ and the ratio of momentum traders $\gamma$= 0.1 (circles), 0.15 (squares). Other parameters are: the total number of individuals $N=1001$, the memory size $m=5$, the updating threshold $D_{th}=0$, the drift of strategy $\lambda=0.1$, the exponent $\xi=1$, the constant $\eta=10$. Final results are averaged over 10 runs and $10^4$ times after $10^5$ relaxation times in each run. }
\end{figure}

In fig. 2 (a) we plot the root mean square fluctuation $F(S)$ of stock returns depending upon the DFA method. It is noted that, compared with the case where there are no momentum traders, an increase in the ratio $\gamma$ of momentum traders leads to an increase in the hurst exponent $h$ of stock returns. As we draw a best fit line to the simulation data, the hurst exponent is found to be $h\sim 0.49061$ for $\gamma=0.1$ and $h\sim 0.53439$ for $\gamma=0.15$. Such a result is in accordance with the empirical findings in refs. \cite{li1, gu1}, where the hurst exponent of stock returns is about $h\sim 0.5$.

In fig. 2 (b) we plot the root mean square fluctuation $F(S)$ of absolute stock returns depending upon the DFA method. The hurst exponent of absolute stock returns is found to be $h\sim 0.75638$ for $\gamma=0.1$ and $h\sim 0.75417$ for $\gamma=0.15$. Such a result is in accordance with the empirical findings in refs. \cite{plerou1, liu1, chen10, qiu1}, where the hurst exponent of stock return volatilities is between $h\sim 0.5$ and $h\sim 1$.

The above simulation results indicate that the characteristics of power-law distribution of stock returns, no memory in stock return and long-range autocorrelation of stock return volatilities can be reproduced in the present model, which implies that the stylized facts in empirical findings could originate from the existence of momentum traders and typical market impact. The effect of heterogeneous investors on price movement has also been discussed in refs. \cite{kukacka1, chen1, chen2, shi1, shi2}. In the present model, the existence of a small amount of momentum traders coupled with variable market impact is in favor of reproducing more diversified financial markets.

\subsection{\label{subsec:levelB} Coevolutionary dynamics of individual strategies and price movement}

In the present work, the self-reinforcing feedback loop is defined as follows. An individual acts according to his subjective perception to the market movement. If his perception is confirmed, his strategy will be imitated by more people and herd behavior occurs. The collective action further confirms the correctness of his subjective perception and his strategy becomes a dominant strategy. Accordingly, the negative feedback loop can also be defined here. An individual acts according to his subjective perception to the market movement. If the market moves contrary to his perception, his strategy will be rejected by other people. The trend-following behavior coexists with the trend-rejecting behavior. The crowd-anticrowd strategy becomes a dominant strategy.

In the following, we firstly investigate the coevolutionary dynamics of individual strategies and price movement and then examine the conditions for the occurrence of self-reinforcing feedback loop in the present model.

\begin{figure}
\includegraphics[width=10cm]{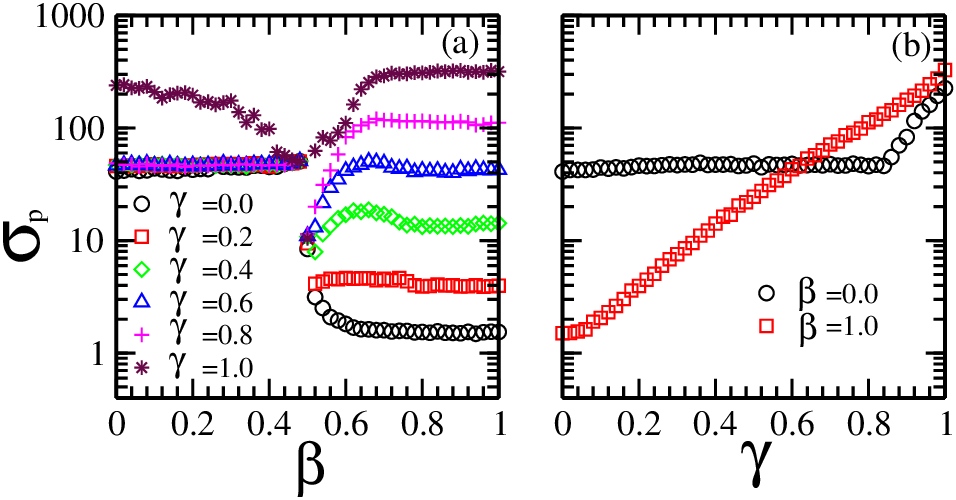}
\caption{\label{fig:epsart}The standard deviation $\sigma_P$ of stock prices (a) as a function of the market impact $\beta$ for the ratio of momentum traders $\gamma$=0 (circles), 0.2 (squares), 0.4 (diamonds), 0.6 (triangles), 0.8 (pluses), 1.0 (stars); (b) as a function of the ratio of momentum traders $\gamma$ for the market impact $\beta$=0 (circles), $\beta$=1.0 (squares). Other parameters are: the total number of individuals $N=1001$, the memory size $m=5$, the updating threshold $D_{th}=0$, the drift of strategy $\lambda=0.1$, the exponent $\xi=1$, the constant $\eta=10$. Final results are averaged over 100 runs and $10^4$ times after $10^5$ relaxation times in each run.}
\end{figure}

Figure 3 (a) shows the standard deviation $\sigma_P$ of stock prices as a function of the market impact $\beta$ for different ratio of momentum traders $\gamma$. As there are no momentum traders, $\sigma_P$ is quite large within the range of $\beta<0.5$ and is relatively small within the range of $\beta>0.5$. A transition point $\beta\sim0.5$ is observed. The dependence of $\sigma_P$ on $\gamma$ is related to $\beta$. For $\beta<0.5$, an increase in $\gamma$ has little impact on $\sigma_P$. For $\beta>0.5$, an increase in $\gamma$ leads to an increase in $\sigma_P$.

Figure 3 (b) shows the dependence of $\sigma_P$ on $\gamma$ for $\beta=0$ and $\beta=1$ respectively. For $\beta=0$, $\sigma_P$ is insensitive to the change of $\gamma$ within the range of $0\le\gamma<0.82$ and increases continuously from $\sigma_P\sim 45$ to $\sigma_P\sim 230$ within the range of $0.82<\gamma\le1$. For $\beta=1$, $\sigma_P$ is sensitive to the change of $\gamma$ and increases continuously from $\sigma_P\sim 1.6$ to $\sigma_P\sim 320$.

\begin{figure}
\includegraphics[width=10cm]{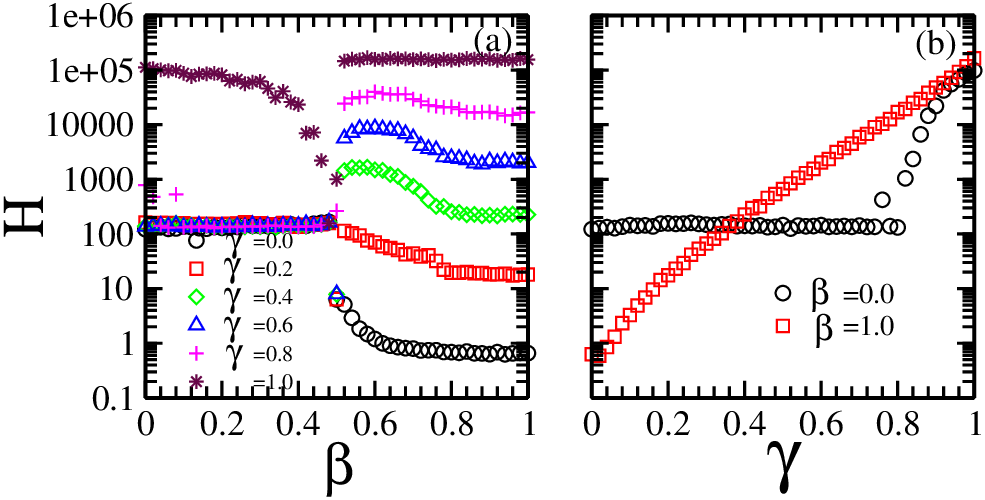}
\caption{\label{fig:epsart} The predictability $H$ of stock prices (a) as a function of the market impact $\beta$ for the ratio of momentum traders $\gamma$=0 (circles), 0.2 (squares), 0.4 (diamonds), 0.6 (triangles), 0.8 (pluses), 1.0 (stars); (b) as a function of the ratio of momentum traders $\gamma$ for the market impact $\beta$=0 (circles) and 1.0 (squares). Other parameters are: the total number of individuals $N=1001$, the memory size $m=5$, the updating threshold $D_{th}=0$, the drift of strategy $\lambda=0.1$, the exponent $\xi=1$, the constant $\eta=10$. Final results are averaged over 100 runs and $10^4$ times after $10^5$ relaxation times in each run.}
\end{figure}

In order to know about whether a large fluctuation in stock prices results from its high predictability, in fig. 4 (a) and (b) we plot the predictability $H$ as a function of the market impact $\beta$ and the ratio of momentum traders $\gamma$ respectively. Similar to the changing tendency of $\sigma_p$ vs $\gamma$ in fig. 3, for $\beta=0$, $H$ is insensitive to the change of $\gamma$ within the range of $0\le\gamma<0.82$ and increases continuously from $H\sim 10^2$ to $H\sim 10^5$ within the range of $0.82<\gamma\le1$. For $\beta=1$, $H$ is sensitive to the change of $\gamma$ and increases continuously from $H\sim 0.6$ to $H\sim 1.5\times10^5$.

Comparing the results in fig. 4 with the results fig. 3, we find that the larger the price fluctuation is, the more predictable the price trend is. 

\begin{figure}
\includegraphics[width=10cm]{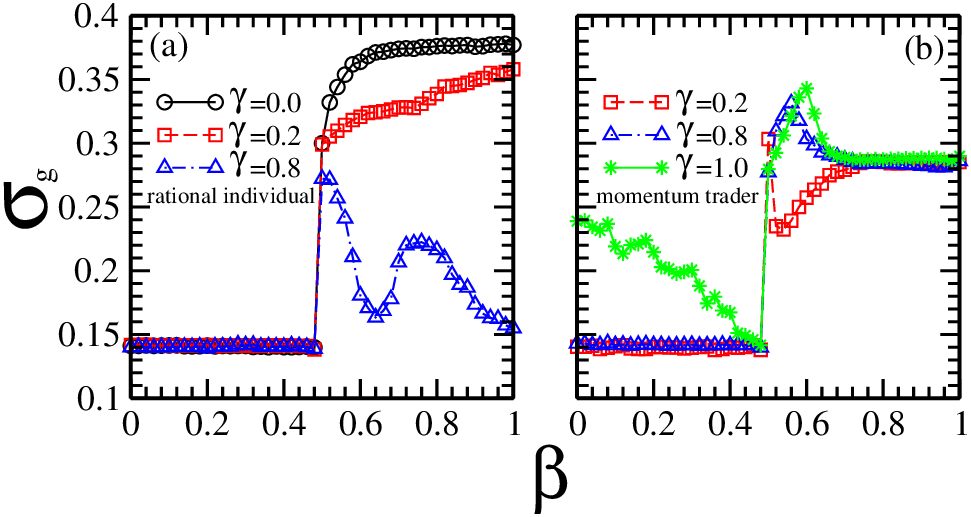}
\caption{\label{fig:epsart}The standard deviation $\sigma_g$ of individual strategies as a function of the market impact $\beta$ for (a) the rational individuals and $\gamma$=0 (circles), 0.2 (squares), 0.8 (triangles); (b) the momentum traders and $\gamma$=0.2 (squares), 0.8 (triangles), 1.0 (stars). Other parameters are: the total number of individuals $N=1001$, the memory size $m=5$, the updating threshold $D_{th}=0$, the drift of strategy $\lambda=0.1$, the exponent $\xi=1$, the constant $\eta=10$. Final results are averaged over 100 runs and $10^4$ times after $10^5$ relaxation times in each run.}
\end{figure}

In the present model, the stock prices coevolve with the strategies. In order to find out the coevolutionary mechanism of individual strategies and stock prices, in fig. 5 we plot the standard deviation $\sigma_g$ of strategies as a function of the market impact $\beta$ for different ratio of momentum traders $\gamma$. It is noted that the evolutionary behaviors of rational individuals and momentum traders are quite different from each other. Within the range of $\beta<0.5$, whether for rational individuals or momentum traders, $\sigma_g$ changes little with the rise of $\gamma$. Within the range of of $\beta>0.5$, an increase in $\gamma$ effectively lowers the dispersion degree of the strategies of rational individuals but has little impact on the dispersion degree of the strategies of momentum traders.

\begin{figure}
\includegraphics[width=12cm]{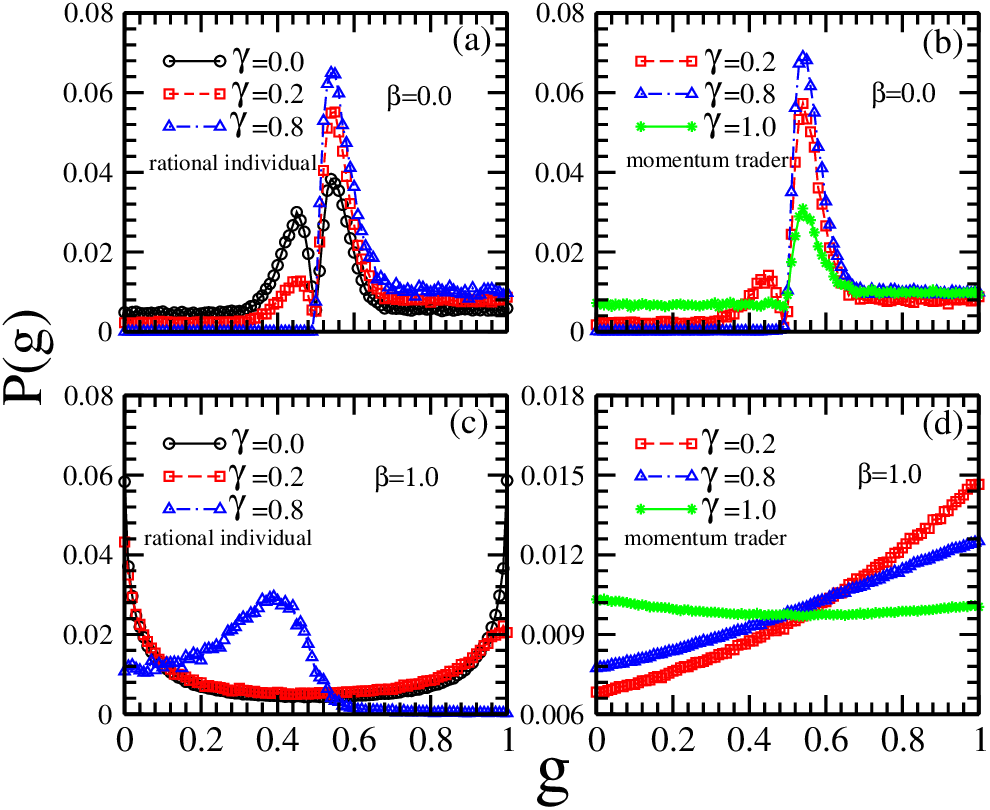}
\caption{\label{fig:epsart} (a) The strategy distribution $P(g)$ of rational individuals for $\beta$=0 and $\gamma$=0 (circles), 0.2 (squares), 0.8 (triangles); (b) the strategy distribution $P(g)$ of momentum traders for $\beta$=0 and $\gamma$=0.2 (squares), 0.8 (triangles), 1.0 (stars); (c) the strategy distribution $P(g)$ of rational individuals for $\beta$=1.0 and $\gamma$=0 (circles), 0.2 (squares), 0.8 (triangles); (d) the strategy distribution $P(g)$ of momentum traders for $\beta$=1.0 and $\gamma$=0.2 (squares), 0.8 (triangles), 1.0 (stars). Other parameters are: the total number of individuals $N=1001$, the memory size $m=5$, the updating threshold $D_{th}=0$, the drift of strategy $\lambda=0.1$, the exponent $\xi=1$, the constant $\eta=10$. Final results are averaged over 100 runs and $10^4$ times after $10^5$ relaxation times in each run.}
\end{figure}

Figure 6 (a) and (b) further show the strategy distributions for $\beta=0$ and $\beta=1$ respectively. It is noted that, for $\beta=0$, the existence of momentum traders leads to an increase in the number of individuals adopting the trend-following strategy, i.e. $g>0.5$. The strategy distributions of rational individuals and momentum traders are the same. For $\beta=1$, the existence of momentum traders suppresses rational individual's trend-following strategy. Rational individual's trend-rejecting strategy coexists with momentum trader's trend-following strategy.

Comparing the results in fig. 3 with the results in fig. 5 and fig. 6, we find that the impact of momentum traders on $\sigma_p$ is closely related to the impact of momentum traders on $\sigma_g$. Within the range of $\gamma<0.5$, an increase in $\gamma$  has little impact on $\sigma_g$ and $\sigma_p$. Within the range of $\gamma>0.5$, an increase in $\gamma$  leads to a decrease in $\sigma_g$ and an increase in $\sigma_p$. Such a result implies that the price fluctuation should be determined by the strategy distribution of rational individuals. A U-like strategy distribution is beneficial to the suppression of price fluctuation. A narrower strategy distribution leads to a large price fluctuation. Both the trend-following and trend-rejecting herd behaviors can result in large price fluctuations. The coexistence of the trend-following and trend-rejecting behaviors can suppress the price fluctuations.

\begin{figure}
\includegraphics[width=10cm]{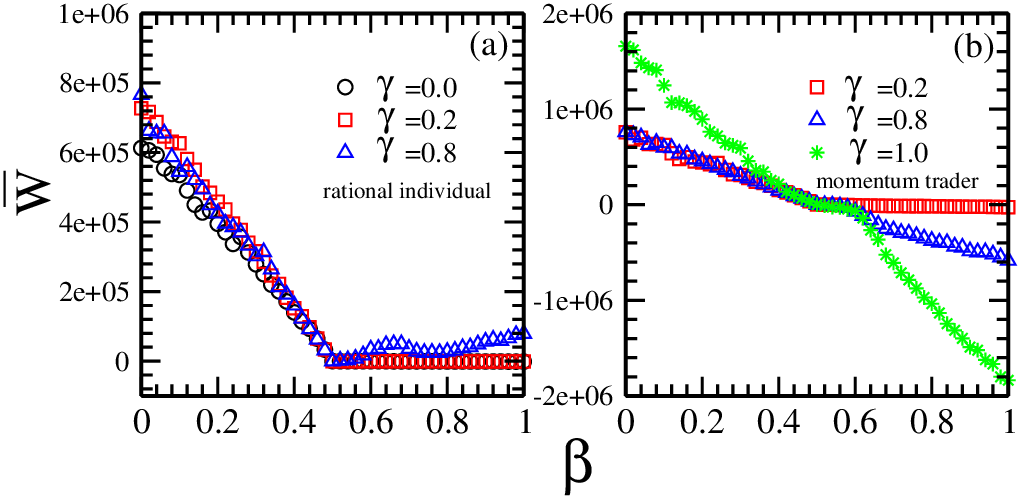}
\caption{\label{fig:epsart}The average wealth $\bar W$ as a function of the market impact $\beta$ for (a) the rational individuals and $\gamma$=0 (circles), 0.2 (squares), 0.8 (triangles); (b) the momentum traders and $\gamma$=0.2 (squares), 0.8 (triangles), 1.0 (stars). Other parameters are: the total number of individuals $N=1001$, the memory size $m=5$, the updating threshold $D_{th}=0$, the drift of strategy $\lambda=0.1$, the exponent $\xi=1$, the constant $\eta=10$. Final results are averaged over 100 runs and $10^4$ times after $10^5$ relaxation times in each run.}
\end{figure}

The evolution of individual strategies is governed by the gains and loses in the investment. In fig. 7 we plot the average wealth $\bar W$ of rational individuals and momentum traders as a function of the market impact $\beta$ for different ratio of momentum traders $\gamma$. Within the range of $\beta<0.5$, the existence of momentum traders leads to an increase in $\bar W$, whether the individuals are rational individuals or momentum traders. Within the range of $\beta>0.5$, the average wealth of momentum traders decreases while the average wealth of rational individual increases with the rise of $\gamma$.

Comparing the results in fig. 6 with those in fig. 7, we find a self-reinforcing feedback loop within the range of $\beta<0.5$. Due to the existence of momentum traders, the trend-followers are more possible to be in the majority side and earn more, which in turn attracts more rational individuals to adopt the trend-following strategy and the trend-following strategy becomes a dominant strategy. However, Within the range of $\beta>0.5$, the above self-reinforcing feedback loop does not exist. On the contrary, a negative feedback loop is found. Due to the existence of momentum traders, the trend-followers are more possible to be in the majority side and suffer a lose, which in turn cause the rational individuals to reject the trend-following strategy and the trend-rejecting strategy becomes a dominant strategy for rational individuals. Due to an increase in the number of rational individuals, the trend-rejectors are more possible to be in the majority side and suffer a loss, which in turn cause the momentum traders to adopt the trend-following strategy and the trend-following strategy becomes a dominant strategy for momentum traders. The rational individuals clustering around $g_i\sim0$ coupled with the momentum traders clustering around $g_i\sim1$ becomes stable.

\section{Theoretical analysis}
\label{sec:analysis}
\subsection{\label{subsec:levelA} The evolutionary dynamics of self-reinforcing feedback loop in the present model}

In the present model, the self-reinforcing feedback loop is observed only when the market impact is low. In the following, we give a theoretical analysis on the self-reinforcing feedback mechanism for $\beta=0$.

Consider the case $\gamma=0$. As what has been discussed in ref. \cite{zhong1}, the evolutionary mechanism is satisfied with the majority game mechanism, where the individuals in the majority side will win the game.

In order to understand the evolutionary mechanism of individual strategies, we can divide them into three categories: $g<0.5$; $g=0.5$; $g>0.5$. If all the individuals adopt the $g<0.5$ strategies, the trend-rejectors should be in the majority side. They will get a net benefit and the $g<0.5$ strategies become stable. If all the individuals adopt the $g>0.5$ strategies, the trend-followers should be in the majority side. They will get a net benefit and the $g>0.5$ strategies become stable. If all the individuals adopt the $g=0.5$ strategy, it is a random case whether the trend-rejectors or the trend-followers are in the majority side. Each one is quite possible to suffer a loss and the $g=0.5$ strategy is unstable. If half of the individuals adopt the $g<0.5$ strategies and another half of the individuals adopt the $g>0.5$ strategies, it is also a random case similar to the case for $g=0.5$. Therefore, the probability that the system evolves to the state where all the individuals adopt the $g<0.5$ strategies is equal to the probability that the system evolves to the state where all the individuals adopt the $g>0.5$ strategies.

As the momentum traders are incorporated into the evolutionary process, the game is also satisfied with the majority game mechanism but the system is more possible to evolve to the state where all the individuals adopt the $g>0.5$ strategies. The difference between the probability that all the individuals adopt the $g>0.5$ strategies and the probability that all the individuals adopt the $g<0.5$ strategies is

\begin{equation}
p'=\sum_{{0\le m',n'\le \frac{N-\gamma N}{2}};m'+n'\ge \frac{N-2\gamma N}{2}}\binom{\frac{N-\gamma N}{2}}{m'}\binom{\frac{N-\gamma N}{2}}{n'}g'^{\frac{N-\gamma N}{2}+m'-n'}(1-g')^{\frac{N-\gamma N}{2}-m'+n'},
\end{equation}
in which $g'$ belongs to $[0,0.5]$ and $m'$ and $n'$ are the numbers of individuals following the trend with probability $g'$ and $1-g'$ respectively. For example, suppose $N=3$ and $\gamma N=1$, $p'$ becomes

\begin{equation}
p'=\sum_{{0\le m',n'\le 1};m'+n'\ge \frac{1}{2}}g'^{1+m'-n'}(1-g')^{1-m'+n'}=g'^2+(1-g')^2+g'(1-g').
\end{equation}

From the above equation we find that the probability that all the individuals adopt the $g>0.5$ strategies increases with the rise of $\gamma$ while the probability that all the individuals adopt the $g<0.5$ strategies decreases with the rise of $\gamma$, which is in accordance with the simulation results in fig. 6 (a) and (b).

\subsection{\label{subsec:levelB} The evolutionary dynamics of negative feedback loop in the present model}

In the present model, the negative feedback loop is observed only when the market impact is high. In the following, we give a theoretical analysis on the negative feedback mechanism for $\beta=1$. 

Consider the case $\gamma=0$. As what has been discussed in ref. \cite{zhong1}, the evolutionary mechanism is satisfied with the minority game mechanism, where the individuals in the minority side will win the game. The functional form of strategy distribution $P(g)$ is determined by the winning probability $\tau(g)$ \cite{lo1},

\begin{equation}
P(g)\propto\frac{1}{\frac{1}{2}-\tau(g)},
\end{equation}
in which $\tau(g)\sim\frac{1}{2}-\frac{1}{\sqrt N}g(1-g)$. For $g = 0$ and $g = 1$, $\tau(g)$ reaches its maximum value $\tau(0)=\tau(1)=\frac{1}{2}$. Therefore, a U-shape distribution clustering around $g = 0$ and $g = 1$ is stable.

As the momentum traders are incorporated into the evolutionary process, the game is also satisfied with the minority game mechanism. If a continuous rise or a continuous fall of the prices does not occur, all the individuals behave like rational individuals. A U-shape distribution clustering around $g = 0$ and $g = 1$ is also stable.

If a continuous rise or a continuous fall of the prices occurs, i.e. $\uparrow\uparrow\uparrow$ or $\downarrow\downarrow\downarrow$ occurs, the $g = 1$ strategy of rational individuals will be suppressed. For example, suppose originally half of the individuals adopt the $g=0$ strategy and another half of the individuals adopt the $g=1$ strategy, among which the ratio of rational individuals is equal to the ratio of momentum traders. Facing the $\uparrow\uparrow\uparrow$ ($\downarrow\downarrow\downarrow$) history, nearly half of the rational individuals buy the stock while another half of the rational individuals sell the stock, but all the momentum traders buys (sells) the stock. The price rises (falls) and the rational individuals following the rising (falling) trend lose the game. The $g=1$ strategy will be updated and the ratio of rational individuals adopting the $g=1$ strategy decreases. Therefore, the rational individuals adopting the $g=0$ strategy should be more than the rational individuals adopting the $g=1$ strategy.

If a continuous rise or a continuous fall of the prices occasionally disappears, the $g = 0$ strategy of momentum traders will be suppressed. For example, suppose originally half of the momentum traders adopt the $g=0$ strategy and another half of the momentum traders adopt the $g=1$ strategy. Facing the latest history, half of the momentum traders buy the stock and another half of the momentum traders sell the stock. Because most of the rational individuals reject the trend, the momentum traders rejecting the trend will be in the majority size and lose the game. The $g=0$ strategy will be updated and the ratio of momentum traders adopting the $g=0$ strategy decreases. Therefore, the momentum traders adopting the $g=1$ strategy should be more than the momentum traders adopting the $g=0$ strategy.

From the above analysis we find that, for $\beta=1$, the system will finally evolve to the state where the rational individuals with a dominant strategy $g=0$ coexist with the momentum traders with a dominant strategy $g=1$, which is in accordance with the simulation results in fig. 6 (c) and (d).

\section{Summary}
\label{sec:summary}
By incorporating market impact and momentum traders into an agent-based model, we have investigated the conditions for the occurrence of self-reinforcing feedback loops in financial markets. We focus on the coupling effect of market impact and  momentum traders on the price fluctuations and the evolution of strategies.

The effect of momentum traders on the evolution of stock prices is closely related to the market impact. As the market impact is low, the existence of momentum traders has little effect on the change of price fluctuations. As the market impact is high, the existence of momentum traders can effectively lead to an increase in price fluctuations.

The self-reinforcing feedback loop only exists within the range where the market impact is low, where the herd behavior is beneficial to the investors. Due to the existence of momentum traders, the trend-followers are more possible to be in the majority side and earn more, which in turn causes the trend-following strategy to become the dominant strategy. Within the range where the market impact is high, the self-reinforcing feedback loop does not exist while the negative feedback loop is observed. Due to the existence of momentum traders, the trend-followers suffer a great lose, which in turn results in such a scenario where the momentum traders adopting the trend-following strategy coexist with the rational individuals adopting the trend-rejecting strategy.

Some empirical findings indicate that the momentum strategy helps the investors earn more while other empirical findings indicate that the momentum strategy causes the investors to suffer a loss. The present work gives a possible explanation for the coexistence of these contradictory situations. Some empirical findings tell us that a rational analysis is quite important for an investor to earn more or reduce the loss. The present model gives a possible explanation for the evolutionary mechanism through which a rational individual can earn more and refrain from getting a loss.

The irrational behaviors widely exist in biological and social systems. Whether they are just a kind of noise or not is a fascinating problem which can help us understand the evolutionary mechanism of social behaviors. In the future, the effect of subjective perception on evolutionary dynamics will be extensively studied in diffusion systems and cooperative systems.

\section*{Acknowledgments}
This work is the research fruits of National Natural Science Foundation of China (Grant Nos. 71371165, 61503109, 71631005, 71273224, 71471161, 71471031, 71171036, 71202039, 71773105), Collegial Laboratory Project of Zhejiang Province (Grant No. YB201628), Humanities and Social Sciences
Fund sponsored by Ministry of Education of China (Grant No. 17YJAZH067),  Zhejiang Provincial Natural Science Foundation of China (Grant No. LY17G030024), Research Project of Generalized Virtual Economy(Grant No. GX2015 -1004(M)), Jiangxi Provincial Young Scientist Training Project (Grant No. 2013 3BCB23017).





\bibliographystyle{model1-num-names}



\end{document}